\documentclass[twoside]{article}

\usepackage{PRIMEarxiv}

\usepackage[utf8]{inputenc} 
\usepackage[T1]{fontenc}    
\usepackage{hyperref}       
\usepackage{url}            
\usepackage{booktabs}       
\usepackage{amsfonts}       
\usepackage{nicefrac}       
\usepackage{microtype}      
\usepackage{lipsum}
\usepackage{fancyhdr}       
\usepackage{graphicx}       
\graphicspath{{media/}}     
\usepackage{natbib}
 \bibpunct[, ]{(}{)}{,}{a}{}{,}%

\usepackage{tikz}
\usetikzlibrary{shapes.geometric, arrows, positioning, fit}
\usetikzlibrary{shapes.symbols}
\usetikzlibrary{arrows.meta}
\usepackage{xcolor}
\usepackage{pgfplots}
\pgfplotsset{compat=1.18}
\usepackage{graphicx}
\graphicspath{{../images/pictures/},{../images/figures/}}
\usepackage{subcaption} 
\usepackage{listings} 
\usepackage{amsmath}
\usepackage{amssymb}
\usepackage{booktabs}
\usepackage{pdflscape}
\usepackage{siunitx}
\usepackage{multirow} 
\usepackage{threeparttable}
\usepackage{rotating}
\title{Dynamic Structures of Knowledge Production: 
Citation Rates in Hydrogen Technologies
}

\author{
  David Dekker \\
  Edinburgh Business School\\ 
  Heriot-Watt University\\
  Edinburgh\\
  \texttt{D.Dekker@hw.ac.uk} \\
  \And
  Dimitirs Christopoulos \\
  Edinburgh Business School\\ 
  Heriot-Watt University /\ Modul University\\
  Edinburgh /\ Vienna\\
  \texttt{D.Christopoulos@hw.ac.uk} \\
   \And
  Heather McGregor \\
  Dubai Campus \\
  Heriot Watt University \\
  Dubai\\
  \texttt{H.McGregor@hw.ac.uk} \\
}

\begin{document}
\definecolor{v_purper}{HTML}{440154}
\definecolor{v_purple}{HTML}{404788}
\definecolor{v_teal}{HTML}{238A8D}
\definecolor{v_grass}{HTML}{55C667}
\definecolor{v_yellow}{HTML}{FDE725}
\definecolor{v_blue}{HTML}{39568C}
\maketitle

\begin{abstract}
    The established relationship between the improvement rate and patent citation network structure of a technological domain suggests a dynamic network model, which we explore in this paper. Because characteristics of the \textit{dynamic} patent citation network structure determine the \textit{constant} improvement rate, this requires a consistent reproduction of that structure over time. These dynamic network models in principle are survival models, which offer flexibility in the specification of multiple levels of variables, such as subdomain effects. Consistent development of citation structure would require proportionality of hazard rates between subdomains development over time. The hazard rate in these models is the probability for a patent to be cited by a new patent given the time since the last citation. We name these 'knowledge production'-models since they capture the rate of output (new patent) given the input (existing patents). The probability of citation of a patent reveals its 'productivity' in terms of how often the patent contributes to new inventions. Some important economic implications follow from these models.  As an illustration, we analyze all patents on the evolving technological frontiers associated to inventions in hydrogen technologies. The knowledge production rates of subdomains in the wider Hydrogen technology domain are distinct, but the proportionality assumption can not be rejected ensuring consistent development, and hence implying constant improvement rates. Our analysis reveals that of the four key technology subdomains, 'production', 'storage', 'distribution', and 'fuel-cells', it are patents in 'distribution' that exhibit the lowest rate of knowledge production. From this we can conclude that 'distribution'-costs will hold a substantial and dominant cost component in the $kg$ price of $H_2$. Further modeling of the dynamic citation structure makes evident that a key network effect (Katz-centrality) predicts the rate of knowledge production. Adding this variable reduces the subdomain effect, which means that this endogenous variable is more informative than the subdomain classification. Also, since the effect is lower in the 'distribution' subdomain compared to the other subdomains, there seems to be an inherent difference in the organization of invention between subdomains. Other results reveal that knowledge acquired by exploitative learning (within subdomain citation) suffers higher (opportunity) costs of patenting. In fact, this could offer an explanation for the slower development of 'distribution'. As the investments needed in this subdomain are usually very high there will be less incentive to reveal and more incentive to monopolize new knowledge.
\end{abstract}


\section*{Introduction}\label{Introduction}
Technological improvements have been widely acknowledge as the main driver of economic growth \citep[e.g.,][]{farmer_how_2016, sahal_theory_1979, schumpeter_capitalism_2005}. Here economic growth is understood to envelop enhanced productivity, cost reductions and efficiency improvements resulting from technological progress \citep[see for an example,][]{moser_patent_2018}. The technological improvement rate (TIR) as an indicator for innovations, and subsequent productivity improvements is subject to competitive dynamics \citep{aghion_competition_2005}, geographical location \citep[e.g.,][]{ter_wal_dynamics_2014}, the importance, recency and immediacy of patents within a technology domain \citep[e.g.,][]{benson_quantitative_2015, benson_improvement_2014}, as well as the modularity and scaling properties of a technology \citep[e.g.,][]{alves_de_campos_technological_2022}. However, these rates are constant implying an exponential growth of technological improvements, albeit different among domains \citep{moore_cramming_1998, benson_quantitative_2015}, these rates are determined by the structure of patent citation networks \citep{benson_quantitative_2015, magee_quantitative_2016, singh_technological_2021}.

Patent data constitute robust information of inventive behavior \citep{griliches_patent_1990, nagaoka_patent_2010}, which reflects the development of knowledge. Especially, scientific and patent citation networks are used to indicate development of knowledge and improvement rates of technology domains \citep[see,][]{hummon_connectivity_1989,benson_framework_2012}. A technology domain is defined as ``...a body of patented inventions achieving the same technological function using the same knowledge and scientific principles." \citep[][p.1]{singh_technological_2021}, which is consistent with \citep[][p.4]{magee_quantitative_2016} and \citep[][p.2]{park_tracing_2017}. There will always remain some ambiguity to patent classification, as they may be essential to multiple domains, and/or be nested in a subdomain which contributes to a more encompassing technological function, while the subdomain's technological function is narrower. Given the definition of technological domains, it is necessary that it also consistently applies to any subdomain \citep[see][]{hannan_concepts_2019}.

In fact, the existence of technological subdomains, or modularity of technology, brings to light a problem that has not been addressed. Namely, that the TIR's in subdomains that make up a domain need not be identical. The consequence is that bottleneck technological subdomains may exists, that become an increasingly larger cost component in the technology domain. Whenever such technology's subdomains are pivotal it may slow down adaption of reliant technologies. Stimulating investments in other subdomains could very well suffer from such bottleneck subdomains. At the same time effectiveness of stimulation policy focused on a subdomain will be disappointing, when expectations are based on domain developments. In hydrogen technology, the Cooperative patent Classification (CPC) distinguishes 4 subdomains, hydrogen storage, hydrogen distribution, hydrogen production, and fuel cells. Our aim in this study is to determine whether there is a bottleneck subdomain for hydrogen technology. 

The commitment to meet the goals in the 2015 Paris climate agreement is contingent on accelerating technological improvement. At the current level of technological development we cannot meet climate goal commitments. Even substantial stimulus packages, such as the Inflation Reduction Act in the U.S.A., are not adequate on their own to achieve sufficient carbon reductions \citep{bistline_emissions_2023}. Therefore the question of whether or not technology domains develop fast enough to sufficiently contribute to these goals, becomes highly relevant to society. Not only to policy makers, but similarly to those that manage and provide resources critical to technological development, such as entrepreneurs, venture capitalists, engineers, universities, and corporate R\&D. Here, we focus on hydrogen technology to provide a demonstration of how the structural analysis of patents can provide unique insights to aid decisions in corporate and public investment.

Success of alternative technology domains in the energy transition, such as hydrogen technology, depends on many factors. In the process of ideation to scaling, decision-making on investment and divestment take central stage as these determine what innovations will grow and which old technologies will be replaced \citep{schumpeter_capitalism_2005, hart_innovation_2005}. Therefore, accurate prediction of performance development, expressed in terms of increased productivity and the rate of technological growth \citep{moore_cramming_1998, wright_factors_1936} in technology domains becomes of crucial importance. It has also been recognized that this is a difficult task and several alternative approaches have been proposed \citep{farmer_how_2016}.

\subsection*{Patent citation networks}
Recent developments in patent citation network research show a strong association between the citation network's structure and technological (performance) improvement rates \citep{benson_framework_2012, benson_quantitative_2015, triulzi_estimating_2020}. These studies demonstrate high explanatory power of exponential growth models based on citation network centrality measures. These models measure characteristics of a domain citation network and relate it to domain level performance measures. Another advantage of the citation network approach is that it allows to study the relative importance of individual patents in a historical context, i.e. after a certain time the relative contribution of patents to the development speed of the domain can be established \citep{alves_de_campos_technological_2022}. However, the approach is limited in assessing the influence of a single new patent, which would require analyses on patent level. We reason that this is in fact necessary. 

An important observation \citep{triulzi_estimating_2020} is that performance improvements are exponential \citep{farmer_how_2016, koh_functional_2006, magee_quantitative_2016, nagy_statistical_2013}. This implies that the rate of improvement is constant \citep{benson_quantitative_2015, magee_quantitative_2016}. Although, the implications of this effect on patent citation network structure have been studied there is another aspect to this observation that deserves attention. As the rate of performance improvement is constant it simultaneously represents a defining function of a \textit{dynamic} network. As patents get added to a domain its citation network could change in many different ways. However in terms of aggregate aspects that influence TIR the network should remain the same for the consistent and strong effects found \citep{benson_quantitative_2015, magee_quantitative_2016}. This motivates the study of technology domains as dynamic patent citation networks, because it suggests structural citation behavior within subdomains. It allows the analysis on the level of patents, where we can still distinguish parameter estimates for different subdomains.

\begin{figure}[t]
  \centering 


\begin{tikzpicture}[node distance=2cm]
  \coordinate (probA) at (0,0);
  \coordinate (probB) at (0,0);
  \coordinate (probC) at (0,0);

  \node[star, star points=5, fill={v_grass}, draw={black}, minimum size=1cm] (starA1) at (5,2.5) {};
  \node[star, star points=5, fill={v_purple}, draw={black}, minimum size=1cm] (starC1) at (-2,2.5) {};

  \node[star, star points=5, fill={v_grass}, draw={black}, minimum size=1cm] (starA2) at (5,-1) {};  
  \node[star, star points=5, fill={v_teal}, draw={black}, minimum size=1cm] (starB1) at (1.5,-1) {};
  \node[star, star points=5, fill={v_purple}, draw={black}, minimum size=1cm] (starC2) at (-2,-1) {};
  \node[star, star points=5, fill={v_grass}, draw={black}, minimum size=1cm] (starA3) at (5,-3.5) {};

  \node[minimum size=2cm] (t0) at (-2.15,-5.3) {$t=0$};
  \node[minimum size=2cm] (t3) at (5.05,-5.3) {$t=3$};

    \path (starA1) edge[-{Latex[length=3mm,width=2mm]}, black!100, thick](starC1) ;
  \path (starA2) edge[-{Latex[length=3mm,width=2mm]}, black!100, thick] (starB1) ;
  \path (starB1) edge[-{Latex[length=3mm,width=2mm]}, black!100, thick] (starC2);
  \path (starA3) edge[-{Latex[length=3mm,width=2mm]}, black!100, thick] node (probC) [draw=black!100, thick, text width=3cm, align=center, left, xshift=0.5cm, yshift=-0.95cm] {Cit. w/in 3 years with $>$50\% probability} (starC2);

  \path (-2.75,-5) edge[-{Latex[length=3mm,width=1mm]}, black!100, ultra thick] node (probC) [draw=black!00, thick, text width=3cm, align=center, midway, yshift=-0.5cm] {\textit{time} $t$ }(6, -5);

  \begin{axis}[
    hide axis,
    scale only axis,
    height=6cm, 
    width=0.5cm, 
    colormap/viridis,
    colorbar,
    colorbar style={
      ylabel style={
        text width=2.5cm, 
        align=center,
        anchor=south,
      },
      title style={align=center}, 
      title={\textbf{Age}},
      ytick={0,1}, 
      yticklabels={Older,Newer}, 
      yticklabel pos=right, 
      at={(15,-0.25)}, 
      anchor=west, 
    }
  ]
    \addplot [draw=none] coordinates {(8,0) (7,1)};
  \end{axis}

  \node[star, star points=5, draw=black, minimum size=1cm, fill=none, above=1.5 cm of current colorbar axis.north] (patentStar) {}; 
  \node[right=0.2cm of patentStar] {\textbf{Patent}};

  \node[draw=v_purper, fit=(starA1) (starC1), inner sep=5mm, label={[above]\textbf{$Domain \, A:$} citations w/in 3 years with $50\%$ probability}] {};
  \node[draw=v_blue, fit=(starA2) (starA3) (starB1) (starC2), inner sep=5mm, label={[above]\textbf{$Domain \, B:$} citations w/in 1.5 years with $50\%$ probability}] {};

\end{tikzpicture}

  \caption{The quantity effect of a higher rate of citation. Over the same period more patents appear that are based on patent at $t=0$. As innovations often focus on enhancing efficiency, more opportunity for cost reductions appear, with the additional increase in the likelihood that these innovations will be unique and difficult to replicate, providing a sustainable competitive cost advantage.}
  \label{fig:citation_rate_quantity_effect}
\end{figure}
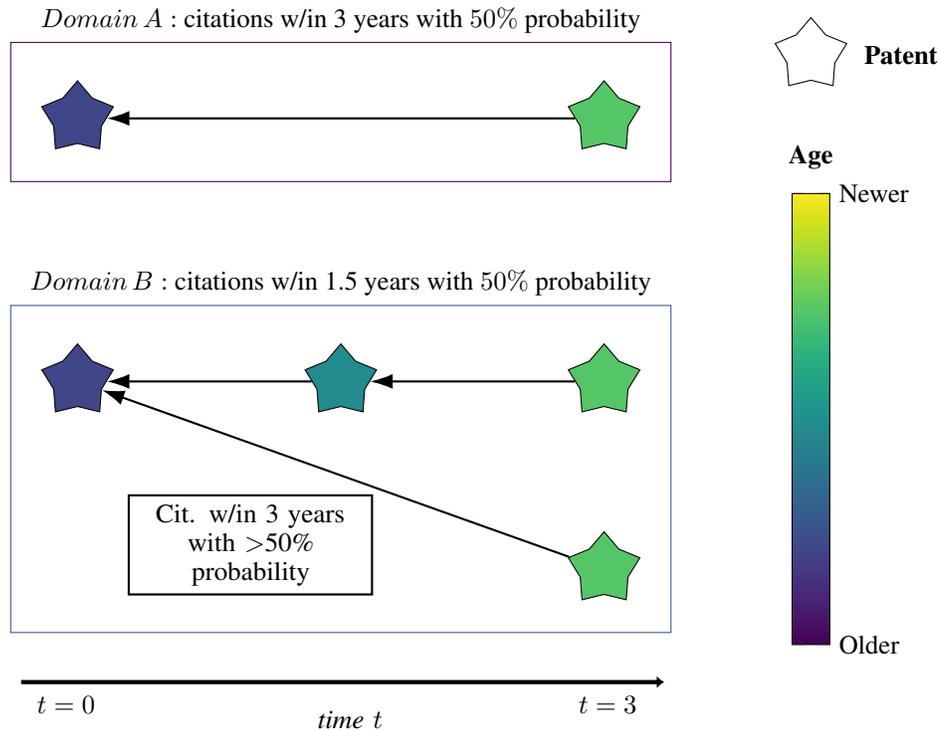

The speed of patent citations as a measure of knowledge production is a function of the dynamic citation network structure, and hence can be interpreted as a direct measure of the technological improvement rate. There are at least two major advantages stemming from looking at citation speed. Comparing the citation speed between subdomains subsequently allows identifying long run proportional weight of cost components \citep[see][on cost functions]{sahal_theory_1979}. To examine the relationship between technological improvement rate and cost components, we focus on the patent citation network as a proxy for innovation and development within a technology domain. We hypothesize that the centrality of patents within the citation network serves as a strong indicator of the citation rate. Furthermore, we argue that changes in citation rate and citation network structure can provide insights into fundamental shifts and directions of development within the domain.

Besides other domain specific characteristics, such as linguistic, geographic, and economic, the citation network directly links micro behavior to aspects attributable to the corpus of patents, and hence the domain it embodies. Furthermore, there are some specific traits of patent citation networks that reveal the (stochastic) relation between citation behavior and the technical improvement rate of the (sub-)domain. Assume we have 2 (sub-)domains $A$ and $B$. In $Domain \, A$ there is a $50\%$ probability that a patent is cited within a period of $3$ years. In $Domain \, B$ this $50\%$ probability is within a period of $1.5$ years. This implies that the domains differ in the productivity of patents in developing new knowledge. If all probabilities are realized, as depicted in Figure \ref{fig:citation_rate_quantity_effect}, after 3 years the productivity of the null-patent ($t=0$) in $Domain \, B$ will be triple of that in $Domain \, A$. This is a simplification of aspects of individual patents. Their embedding in the patent network will affect this probability at given time points. However, the example illustrates that a higher citation rate over a given period, implies a higher knowledge production rate in the subdomain. Consequently, the likelihood that cost advantages occur increases, at an accelerating rate. We coin this the quantity effect of patent citation rates, which explains a higher TIR due to increase in the average recency \citep[see][]{singh_technological_2021}.  It can be seen to reflect the dynamism and potential for growth within a specific technological domain and can be compared among competing or complementary technologies.

\subsection*{Hydrogen technology}
The prediction of cost components in technology domains, such as production, storage, and distribution, is crucial for effective resource allocation and decision-making. However, accurately forecasting cost levels in these domains is challenging due to various factors and uncertainties. By implication we assume that the relative cost component ratio is proportional to the TIR ratio between subdomains, which can be derived from the citation rates within different (sub-)domains.

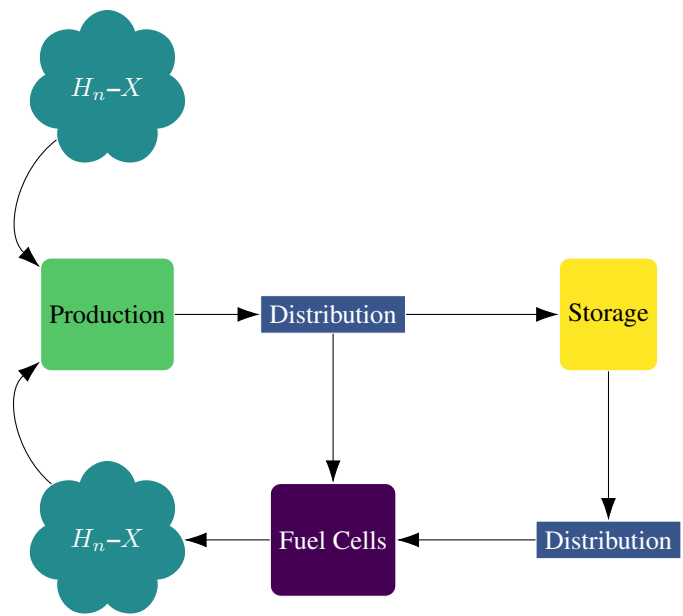
\begin{figure}[t]
  \centering 

\begin{tikzpicture}[node distance=3cm]
\tikzstyle{production} = [rectangle, rounded corners, draw=white, fill={v_grass}, text centered, minimum height=1.5cm]
\tikzstyle{distribution} = [rectangle, draw=white, text=white, fill=v_blue, text centered, minimum height=0.5cm]
\tikzstyle{storage} = [rectangle, rounded corners, draw=white, fill={v_yellow}, text centered, minimum height=1.5cm]
\tikzstyle{fuel_cells} = [rectangle, rounded corners, draw=white, text=white, fill={v_purper}, text centered, minimum height=1.5cm]
\tikzstyle{water} = [cloud, cloud puffs=7, draw=white, fill={v_teal}, text=white, minimum size=1.5cm]

\node (water1) [water, left] {$H_n$\textbf{--}$X$};
\node (production) [production, below of=water1] {Production};
\node (distribution1) [distribution, right of=production] {Distribution};
\node (storage) [storage, right of=distribution1, right] {Storage};
\node (distribution2) [distribution, below of=storage] {Distribution};
\node (use) [fuel_cells, below of=distribution1] {Fuel Cells};
\node (water) [water, left of=use] {$H_n$\textbf{--}$X$};

\path (water1) edge[bend right=30, out=-45, in=-125, -{Latex[length=3mm,width=2mm]}] (production);
\path (production) edge[-{Latex[length=3mm,width=2mm]}] (distribution1);
\path (distribution1) edge[-{Latex[length=3mm,width=2mm]}] (storage);
\path (distribution1) edge[-{Latex[length=3mm,width=2mm]}] (use);
\path (storage) edge[-{Latex[length=3mm,width=2mm]}] (distribution2);
\path (distribution2) edge[-{Latex[length=3mm,width=2mm]}] (use);
\path (use) edge[-{Latex[length=3mm,width=2mm]}] (water);
\path (water) edge[bend left=30, out=45, in=125, -{Latex[length=3mm,width=2mm]}] (production);
\end{tikzpicture}
  \caption{Operational relations between hydrogen technology subdomains. These are hydrogen production (non-carbon sources), -storage, -usage (in fuel cells) and -distribution Cooperative Patent Classification (CPC) of Hydrogen technology [codes Y02E60/30: 32, 34, 36, 50]. }
  \label{fig:hydrogen_process}
\end{figure}

We can however already see from Figure \ref{fig:hydrogen_process} that the hydrogen distribution subdomain is an enabling technology  \citep{bresnahan_general_1995, gambardella_profiting_2021}, i.e. a pivotal technology for the other three subdomains. In fact, it could be deemed doubtful that any substantial market for hydrogen could ever exist if distribution costs are not equivalent to substitutes and presumably marginal to the overall cost of hydrogen. If it cannot be distributed Hydrogen will otherwise be restricted to use in proximity to production facilities.

\section*{Data and methods}
\subsection*{Patent citations}
Patent data provide a reliable means to measure different aspects of   both individual and collective economic behaviors \citep[e.g.,][]{griliches_patent_1990, nagaoka_patent_2010, kogan_technological_2017}. Patent citation data specifically has been shown to be informative about the importance of individual patents, future corporate R\&D activity and stock market valuation \citep{trajtenberg_penny_1990, hall_market_2005, fronzetti_colladon_new_2025}, as well as highly associated to domain level technological improvement rates \citep{benson_quantitative_2015, magee_quantitative_2016, triulzi_estimating_2020, singh_technological_2021}.

We retrieve data from "Lens.org" \citep{jefferson_comment_2017}, a literature and patents database, on all issued patents ever in the CPC category "reduction of greenhouse gas (GHG) emissions, related to energy generation, transmission or distribution (Y02E), in the subclass "Enabling technologies: Technologies with a potential or indirect contribution to GHG emissions mitigation" (60), specifically "Hydrogen Storage" (60/32), "Hydrogen distribution" (60/34), "Hydrogen production from non-carbon containing sources, e.g. by water electrolysis" (60/36), and "Fuel cells" (60/50) (see Appendix \ref{appendix:API} for further details on data retrieval). 

\begin{figure}[ht]
\centering
  \includegraphics[bb=0 0 800 600, width=\linewidth]{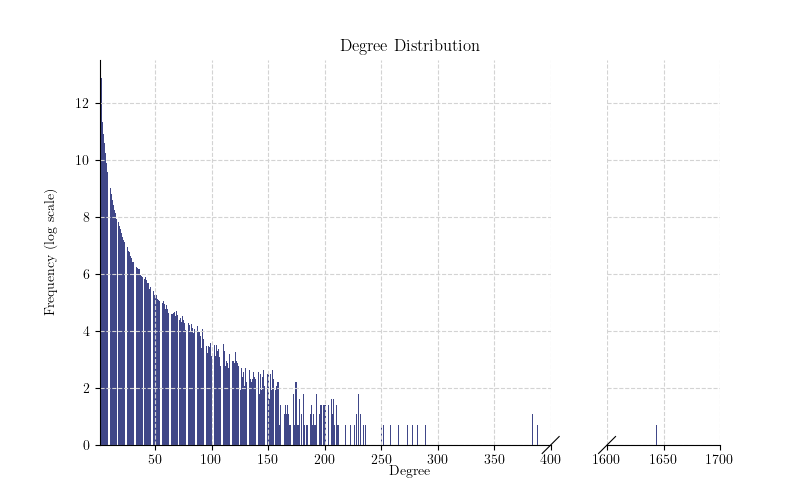} 
    \caption{Degree distribution of patent citation network in $H_{2}$-technology domain.}
\label{fig:degree_dist_hydrogen_tech}
\end{figure}

\begin{figure}[ht]
\centering
  \includegraphics[width=\linewidth]{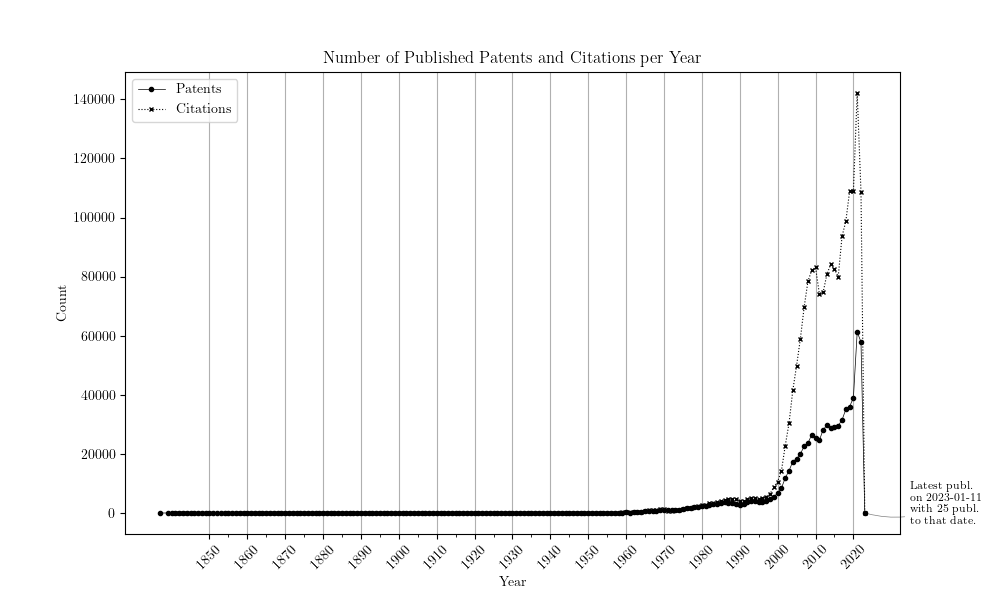} 
    \caption{Number of patents published and citations yearly in the Hydrogen technology domain (CPC Y02E 60).}
\label{fig:publications_hydrogen_tech}
\end{figure}

In total there are $777,695$ patents that receive citations ranging from $1$ to $1660$ (see Figure\ref{fig:degree_dist_hydrogen_tech}). The earliest patent was issued in $1841$ and the latest patent was issued in 2023, spanning a period of 167 years. In Figure \ref{fig:publications_hydrogen_tech} we show the yearly development of the number of patents published and citations in "Hydrogen Technology" (Y02E 60). It is clear that both the number of patents as well as the citations have exploded since the late 1990's, although this growth dipped after 2010, and only recently shows signs of growth again. From Table \ref{tab:patents_descriptives} we see that $50\%$ of patents in this study are published after August 2013.    
\begin{table}
    \centering
        \caption{Descriptive Statistics Patents}
        \begin{tabular}{lrrrrr}
            \toprule
                {} &   Mean &  Median &    Std &    Max &    Min \\
                \midrule

                $H_{2}$ Storage   &  0.044 &  0.000 &  0.291 &    1.000 &  0.000 \\
                $H_{2}$ Distribution &  0.002 &  0.000 &  0.061 &    1.000 &  0.000 \\
                $H_{2}$ Production &  0.065 &  0.000 &  0.343 &    1.000 &  0.000 \\
                Fuel cells        &  0.376 &  1.000 &  0.408 &    1.000 &  0.000 \\
                \midrule
                Pub. Date  &  2010-05-28 &  2013-08-07 &      &  2023-01-06 &  1841-08-04 \\
                Tot. class (CPC) &  8.106 &  6.000 &  8.511 &  166.000 &  1.000 \\
                In-Degree & 1.700 & 0 & 5.922& 1011& 0\\
                Out-Degree & 1.700 & 0 & 7.950& 1660& 0\\
            \bottomrule
        \end{tabular}
    \label{tab:patents_descriptives}
\end{table}

\begin{sidewaysfigure}
\centering
\begin{subfigure}{0.45\textwidth} 
  \centering
  \includegraphics[width=\linewidth]{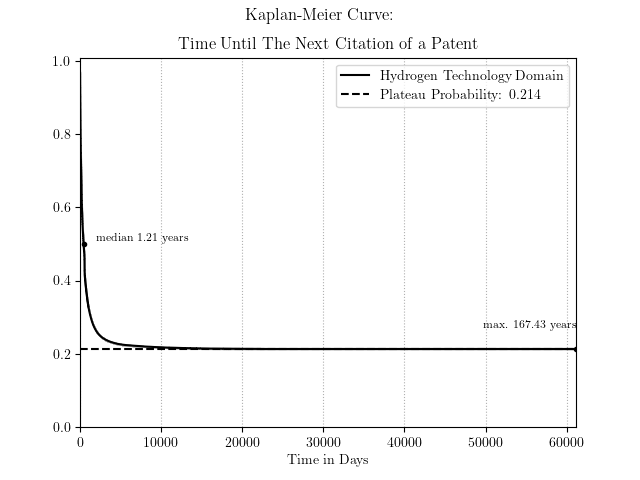} 
  \caption{Hydrogen domain}
  \label{subfig:Hydrogen_domain}
\end{subfigure}
\hfill 
\begin{subfigure}{0.45\textwidth} 
  \centering
  \includegraphics[width=\linewidth]{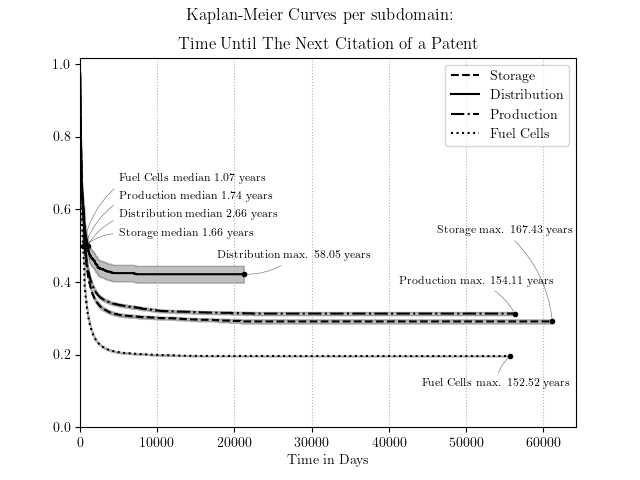} 
  \caption{Hydrogen Subdomains}
  \label{subfig:Hydrogen_Subdomains}
\end{subfigure}
\caption{Citation rates in hydrogen technology over 200 years. The median years to next citation are $1.21$ over the whole domain, $1.66$ (hydrogen storage), $2.66$ (hydrogen distribution), $1.74$ (hydrogen production from non-carbon sources), $1.07$ (hydrogen usage in fuel cells). These concentrate around the left of the graphs, where the drop probability is largest, and most impact full. }
\label{fig:citation_rates_hydrogen}
\end{sidewaysfigure}

\begin{table}
    \centering
    \caption{Descriptive statistics of the directed acyclic patent citation graph from 1841 to 2023, recording 1,791,203 events (1,314,970 citations of 979,212 occur within subdomains) and non-events (476,233 possible next citations didn't occur before end of observed period) in the Hydrogen Technology Domain. }
    \label{tab:descriptive_stats_GRAPH}
\begin{tabular}{lcccc}
\toprule
 & \multicolumn{4}{c}{Time} \\
\cmidrule(lr){2-5}
 & Mean & St.Dev. & Med. & Max. \\
\midrule
Days aft. prev. event & 1563.071 & 3274.400 & 364.000 & 66269.000 \\
Prev. event date & 2011-01-07 & --- & 2012-08-02 & 2022-12-27 \\
Event date & 2012-06-09 & --- & 2014-02-20 & 2023-01-11 \\
\midrule
 & \multicolumn{4}{c}{Patent} \\
\cmidrule(lr){2-5}
 & Mean & St.Dev. & Med. & Max. \\
\midrule
$H_{2}$ Storage & 0.037 & 0.188 & 0.000 & 1.000 \\
$H_{2}$ Distribution & 0.001 & 0.035 & 0.000 & 1.000 \\
$H_{2}$ Production & 0.050 & 0.219 & 0.000 & 1.000 \\
Fuel cells & 0.414 & 0.493 & 0.000 & 1.000 \\
Tot. CPC classes & 8.959 & 9.260 & 7.000 & 166.000 \\
Patents Katz C. & 0.001 & 0.001 & 0.001 & 0.032 \\
\midrule
 & \multicolumn{4}{c}{Citation} \\
\cmidrule(lr){2-5}
 & Mean & St.Dev. & Med. & Max. \\
\midrule
Prop. citation events & 0.734 & 0.440 & 1.000 & 1.000 \\
Shared subdomain & 0.453 & 0.498 & 0.000 & 1.000 \\
\bottomrule
\end{tabular}
\end{table}
From the Kaplan-Meier curves in Figure \ref{fig:citation_rates_hydrogen} we can see qualitative differences between the subdomains. Especially hydrogen distribution is a sub-domain that develops at a slower pace. The probability that an event (next citation) has not occurred for patents in this subdomain has a much higher median value ($2.66$ years) compared to the whole domain ($1.21$ years) the other sub-domains ($1.07$, $1.74$, $1.66$ years for Fuel Cells, Production, and Storage, respectively). This means that the 50\% probability of a next citation in hydrogen distribution takes $2.49$ times longer to reach than in Fuel Cells. The implication for the expected quantity effect can be seen from the example in Figure \ref{fig:citation_rate_quantity_effect}.

Patent citation data are inherently a directed acyclic graph (DAG). As exemplified in Figure \ref{fig:citation_rate_quantity_effect} a patent's backward citation reveals an input-output identifying the source of knowledge accumulation. As patent citation networks evolve over time theoretically no cycles can occur as a patent cannot cite a future patent. In practice there are some rare events where this might occur for patents, when publication dates are equal. In those case we applied a second ordering based on filing date, which took care of reciprocal ties (minimal cycles). Even less cycles with a longer length where identified. Due to later applications being published sooner. In those cases we deleted the citation of the first patent to the newer patent. We checked, but could not detect any effects of this on the outcomes.

 Representing patent citation as a DAG allows to model knowledge production paths and the rate at which these paths develop. At any moment in time a patent citation network can be observed that incorporates all patent citations of previous periods up to that moment. This implies that network descriptives of patents are independent of future network additions, while they are dependent on all past citations. This implies a temporal causality \citep{an_landscape_2018}, which suggests the survival analysis approach is optimal for  capturing the dynamic development of knowledge production within patent subdomains.

\subsection*{Relational event models}
The structure of patent citation networks has a self-referencing effect on the citations a patent receives (forward citations). For example, an inclination towards transitivity (citing references of references) has been found \citep{an_landscape_2018, chakraborty_patent_2020}, while some patents are more important than other patents in the development of a domain \citep{fronzetti_colladon_new_2025}. These effects have been found with the usage of exponential random graph models (ERGM), which where specifically developed to estimates autocorrelation effects of network structures \citep{holland_exponential_1981, frank_markov_1986, wasserman_logit_1996, snijders_markov_2002, hunter_inference_2006}. Subsequently, multiple other models have been developed to model network processes.   

Benson and Magee \citep{benson_quantitative_2015} identify three hypothesized concepts, which can impact the technological improvement rate of domains. They find strong relations between measures of "Importance of Patents",  "Recency" and, "Immediacy" and the independently estimated, domain specific improvement rate coefficients of an exponential growth model. While they propose a range of different measures the best fitting and most parsimonious model ($R^{2} = 0.64$) entails only 2 variables; the average publication year for all patents in a domain (as a proxy of recency) and the average number of citations that a domain patent receives within 3 years of publication (as a joint measure of immediacy and importance). Their results further motivate our use of a relational event model, as they constitute time ratio variables. This implies that over time  values for each patent will change. Furthermore, given that these are effects on constant rate parameters, they should be proportional over time.

Here our interest is with the rate of knowledge development as can be inferred from the citations a patent receives. The underlying assumption is that the knowledge a patent embodies is a building block, an input in the creation of new knowledge. Than the productivity of a patent in producing knowledge becomes higher as the time between subsequent citations becomes shorter, i.e. the citation rate goes up. In other words, this is captured by immediacy which hence is a conditional variable in the rate of knowledge production. 

A common way to estimate rates is with survival models, which estimate the probability of events after a certain amount of time has passed. We define a citation as the event of interest, and measure the time to the next citation a patent receives. A citation signals the production of new knowledge.

The particularity in dynamic patent citation data, which is not dealt with in traditional survival models, is that there is an inherent dependence between patents due to the citations. Relational Event Models (REM) extend the use of survival models into the context of network data. These models mainly focus on when relational events occur and the likelihoods of sequences of events, given their interdependence in the network. In contrast, traditional survival models usually focus on the time to event, and the probability of events occurring, assuming the independence of events. Here we want to estimate just that; the probability of a patent citation, given the time since its last citation, depending on the importance of the patent in the citation network.

As mentioned patent citation data represent a specific type of network (DAG), which implies that existing patents can only receive citations. Hence this meets the Markov assumption in REM's, which states that events are only dependent on the current state of the network \citep{butts_4_2008}. Another advantage is that it also simplifies the event dependency structure as only activity of a new patent needs to be considered, while the past events (patents' addition to the event set and citations) will not change. Hence, we can distinguish between $2$ sets of patents at any time $t$, those that can cite $\mathcal{S}$, and a set of patents that can only be cited $\mathcal{R}$. Further $\mathcal{C} \subseteq \mathbf{c}$, i.e. only one type of event occurs. In technical terms this implies the support set, consisting of all possible events, $\mathbb{A}(\mathcal{A}_t) \subseteq \mathcal{S}_t \times \mathcal{R}_t \times \mathcal{C}$, where $\mathcal{S}_t \cap \mathcal{R}_t = \emptyset$, while  $\mathcal{R}_{t-1} \cup  \mathcal{S}_{t-1} = \mathcal{R}_{t}$ \citep[see][for notations and likelihood functions]{butts_4_2008}. 

Now, when a next citation has \textit{not} occurred at time $t=\tau$, the hazard function, $h(t=\tau)$, gives the citation rate at instance $\tau$, and the survival function $P(c_{ijt}| \mathcal{S}_t, \mathcal{R}_t, t>\tau)$ gives the probability that a patent does not receive a next citation. The relation between the hazard function and survival function is,

\begin{equation}\label{eq:hazard_gen}
    h(c_t) = \frac{f(c_t)}{S(c_t)} = -\frac{d}{dt} \ln{S(c_t)}.
\end{equation}

Now recall that the technological improvement rate was found to be constant and a function of the citation network structure \citep{triulzi_looking_2015, benson_quantitative_2015, singh_technological_2021}. An obvious assumption is that the citation network is self-replicating its structure over time, and hence that the citation rate in Eq. (\ref{eq:hazard_gen}) is a function endogenous and exogenous factors, $h(c_t)=\lambda(i(c_t), j(c_t), \tau(c_t), \mathbf{x_{i,j,t}} )$, $\mathbf{x_{i,j,t}}$ is a vector of exogenous explanatory variables for the citation of patent $j$ by patent $i$ at moment $t=\tau$.

The Cox proportional hazards model is a common model to estimate the hazard function.  
$$h_{ij}(t) = h_0(t) \exp(\alpha \mathbf{x}_{ij} + \beta \mathbf{w}_{ij} + \gamma \mathbf{z}_{ij,d})$$
Where $ h_0(t)$ is the baseline hazard function, $ \alpha$, $\beta$, and $\gamma$ are vectors of coefficients for endogenous ($x_{ij}$),  exogenous patent level variables ($z_{ij}$), and, exogenous domain specific ($w_{ij,d}$),  respectively. Note, that the effects of the explanatory variables are assumed to be multiplicative on the hazard.

\subsubsection*{Exogenous effects}
The exogenous patent level variables are due to the classification of the patent bureau and inventors. By definitions patents refer to newly developed technology, which often can be applied in a number of domains, or even may, in time, spawn their own domain. The breath of application of a patent could imply that it becomes a citation source. This also depends on whether the patent is filed into other subdomains, or other domains all together. We distinguish between the total number of classes (within and beyond the $H_2$ technology domain) a patent is filed in to capture the breadth to which knowledge can be applied. As such patents with a broad range will serve a qualitative difference in comparison to those with a narrow range of domains. A distinctions which parallels that between exploratory and exploitative learning made in the organizational behavior literature \citep{march_exploration_1991}. Exploitative learning will primarily increase efficiency of existing processes, while exploratory learning contributes to creation of new methods or products \citep{march_exploration_1991}. This effect could also be indirect through the cited patents, for which we take again the number of total classes of the patents that a patent cites. Exploitative learning will be characterized by more within domain citations. Therefore, we add a qualitative variable that classifies whether the citing patent at least shares one $H_2$ technology \textit{subdomain} with the cited patent. Note that these are time invariant measures determined at time of publication.

We estimate the hazard rates at domain level, which allows us to see if there is a difference between the proportional hazards in these models. More specifically, we take the "Hydrogen distribution"-domain as a baseline.  

\subsubsection*{Endogenous effects}
Importance of patents as measured by average degree centrality affects technology improvement rate \citep{benson_quantitative_2015}. In \citep{fronzetti_colladon_new_2025} the authors include a composite of centrality measures in mapping technological independence of patents, including Katz centrality \citep{katz_new_1953}, degree centrality, betweennness centrality, closeness centrality \citep{freeman_set_1977} and distinctiveness \citep{fronzetti_colladon_distinctiveness_2020}. We do see value in that approach when assessing technological independence, yet in our model it would introduce ambiguity. We want to capture dependence of citations. We choose Katz-centrality as within a DAG this unambiguously and consistently captures short and long term importance, without consideration for speed of citations. Especially, since we control separately for patent age measure captures the structural importance of a patent independent of recency.   

An advantage of relational event models for patent citations is that they model citations as a function of endogenous structural characteristics of patents in the network. Katz centrality measures nodal importance in a network as it takes into account both direct and indirect connections. It can identify influential patents that may not have a high number of direct citations, but are cited by other highly cited patents. This captures the indirect influence of a patented invention on subsequent inventions. Furthermore, in a DAG, Katz centrality can trace how knowledge propagates through the network. Patents with high Katz centrality are not only important sources of knowledge but also act as key conduits for disseminating knowledge to subsequent inventions, which allows to predict paths of knowledge development.

Katz centrality of node $i$ at time $t$:

\begin{equation*}
    C_{Katz_{t}}(i) = \sum_{k=1}^{\infty} \sum_{j=1}^{n} \alpha^k (A^k)_{ij,t}
\end{equation*}

where, $A$ is the adjacency matrix of the network, $\alpha$ is the attenuation factor and $n$ is the number of nodes in the network. Note a limit on the size of the attenuation factor is $\alpha < {1}/{\lambda_{max}}$ \citep{foster_no_2001}. Given that Katz-centrality is calculated on a DAG, as emphasized in \citep{katz_new_1953}, Gershgorin's Circle Theorem ($\lambda_{max}>max(ID,OD)$) provides the sufficient condition, $\alpha = {1}/{max(Id, OD)+1}$. We establish the Katz-centrality of existing patents on a yearly bases since their publication, which provides sufficiently fine-grained data, while keeping the calculations on our large dataset within finite time. Calculations where done with the \citep{foster_no_2001} implementation in the 'cugraph'-package in python \citep{rees_katz_2023} in combination with the `networkx'-package \citep{hagberg_exploring_2008}.

Also, based on earlier findings patent age matters, which in \citep{benson_quantitative_2015} is conceptualized as "recency". They find a significant effect of recency on technology improvement rates. Age is expected to be related to the time of next citation, because as knowledge develops, older patents are supposed to become less relevant and knowledge becomes "common" and embedded in newer patents \citep[cf.][]{dimaggio_iron_1983, cohen_absorptive_1990}.

\subsection*{Tests}
From the Kaplan-Meier curves in Figure \ref{subfig:Hydrogen_Subdomains} a stark difference in citation rates or knowledge production rates could be inferred. Standard z-test will be used to asses statistical significance of the model parameters.

We estimate the models both with clustering on cited patents and use robust estimators to remedy correlation between observations (patents can be cited multiple times) as well as potential heteroskedasticity due to misspecification. We compare four models, which incorporate different subsets of explanatory variables. We use a likelihood ratio test to test to what extend these models improve.

To asses proportionality we perform a Schoenfeld Residuals Test to test the proportionality assumption underlying the CPH-model. No evidence against the proportionality assumption is found. This helps understanding the self-replicating function of the citation network. Together with the vanishing fixed effects of some subdomain dummies it implies that the covariates explain much of the differences in the rate of development of subdomains. Especially, since  structural variables replace the subdomain effects, this shows that the subdomain can be distinguished by their network structural characteristics.

Fit of these models is more difficult to assess, since we have a highly skewed indegree distribution \citep[see also][]{hung_examining_2010}, as well as many censored observations (not cited patents), which skews the typical concordance measure towards $0.5$, rendering this measure uninformative. This is not problematic as we study the systematic distinction in citation rates between subdomains. The difference is demonstrated in the KM curves, while proportionality holds in the models, showing the effect is systematic.


\section*{Results}
\begin{sidewaysfigure}
        \centering
        \begin{subfigure}{0.45\textwidth} 
          \centering
          \includegraphics[width=\linewidth]{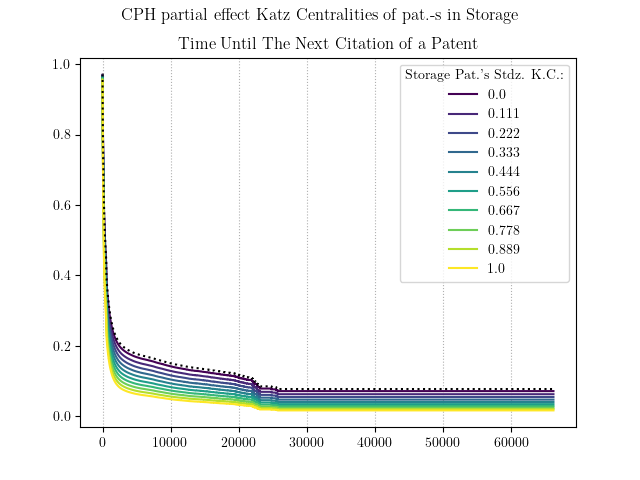} 
          \caption{Partial effect Patents' Katz Centrality in Storage Technology subdomain}
          \label{subfig:Hydrogen_Subdomains_Storage_Partial}
        \end{subfigure}
        \hfill 
        \begin{subfigure}{0.45\textwidth} 
          \centering
          \includegraphics[width=\linewidth]{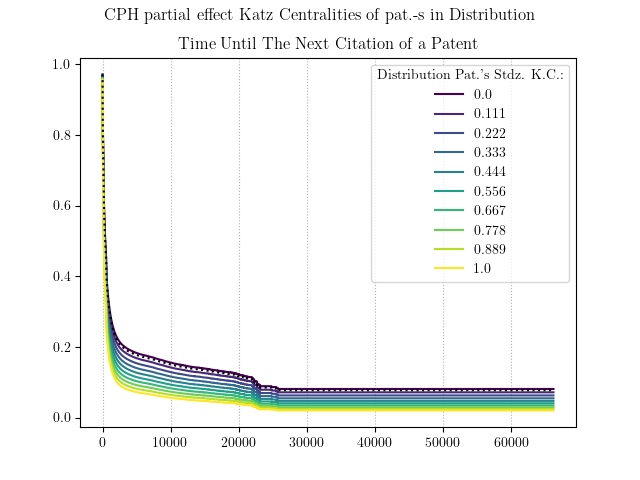} 
          \caption{Partial effect Patents' Katz Centrality in Distribution Technology subdomain}
          \label{subfig:Hydrogen_Subdomains_Distribution_Partial}
        \end{subfigure}
        
        \vspace{0.5cm}
        
        \begin{subfigure}{0.45\textwidth} 
          \centering
          \includegraphics[width=\linewidth]{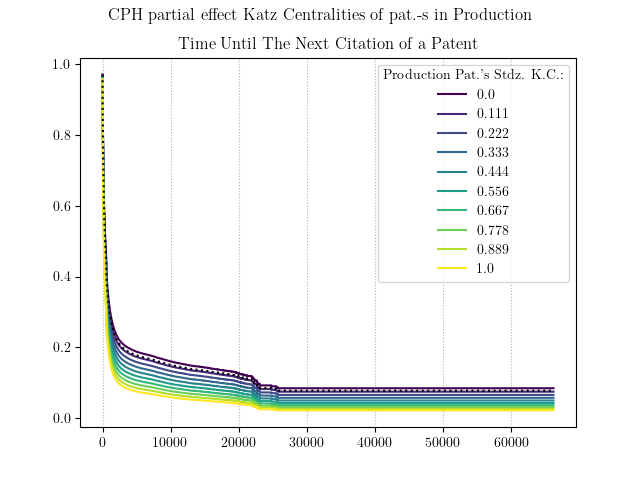} 
          \caption{Partial effect Patents' Katz Centrality in Production Technology subdomain}
          \label{subfig:Hydrogen_Subdomains_Production_Partial}
        \end{subfigure}
        \hfill 
        \begin{subfigure}{0.45\textwidth} 
          \centering
          \includegraphics[width=\linewidth]{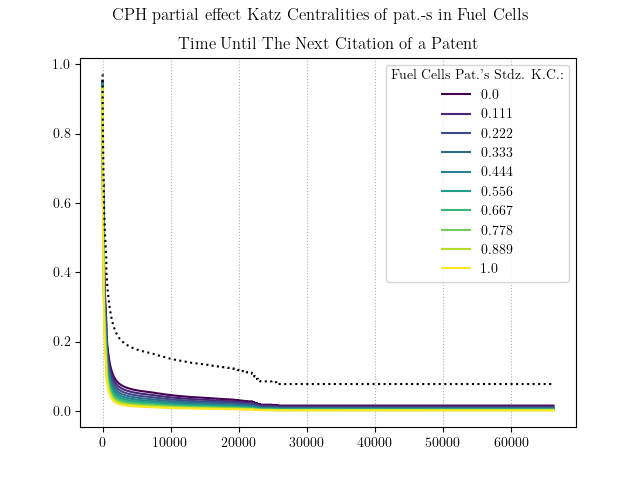} 
          \caption{Partial effect Patents' Katz Centrality in Fuel Cell Technology subdomain}
          \label{subfig:Hydrogen_Subdomains_Fuel_Cells_Partial}
        \end{subfigure}
        \caption{ }
        \label{fig:citation_rates_hydrogen2}
\end{sidewaysfigure}

Our analysis reveals differential citation rates between domains, indicating variations in technological improvement rates as this systematically affects the measures used in \citep{benson_quantitative_2015, magee_quantitative_2016}. The Kaplan-Meier curves in Figure \ref{fig:citation_rates_hydrogen2} show that the median citation over the whole domain is $1.21$ years, this is $2.66 $ for hydrogen distribution, more that twice as long. Furthermore, the other domains show a $1.5$, $1.6$ to about $2.5$ times faster median knowledge production rate, respectively for hydrogen production from non-carbon sources, hydrogen storage, and hydrogen usage in fuel cells. We observe that citation rates are highly dependent on the citation network structure, particularly the centrality of patents, as well as on the age of patents. Older patents have a lower probability of receiving a next citation. And, also the probability of not being cited for the whole domain increases over time. The effect that renders older inventions obsolete, indicates that knowledge develops. That this differs per subdomain indicates distinctive rates of knowledge production. In slower developing subdomains old patents ($>50$ years old), still hold a reasonable probability for a next citation. This signals a reliance on old inventions, which indeed in earlier findings is found characteristic for domains with a low technological improvement rate \citep[e.g.,][]{benson_quantitative_2015, magee_quantitative_2016}.

The differences between the subdomains are substantial and significant. A central finding is that the citation structure self-replicates over time, suggesting the emergence of structured inventive behaviors within technology domains. We see that the proportional hazard test on the null that the hazard ratio between the subdomains is constant, cannot be rejected. This suggests the subdomain effect is consistent over time and in line with the hazard model assumptions. 

\begin{figure}[ht]
\centering
  \centering
  \includegraphics[width=\linewidth]{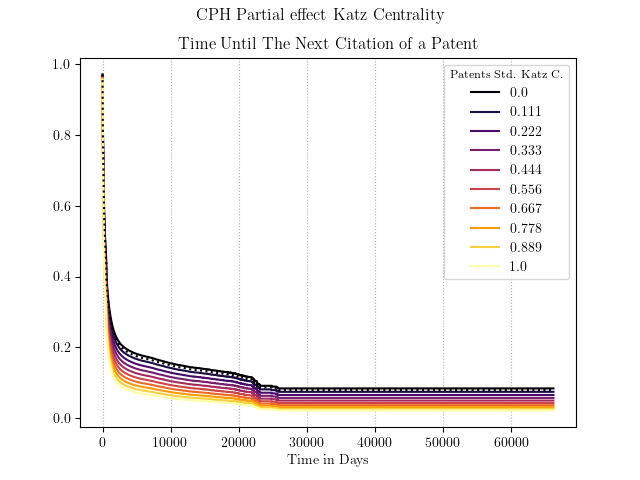} 
\vspace{1cm}
  \caption{Partial effect Patents' Katz Centrality in Hydrogen Technology domain}
  \label{fig:Hydrogen_domain_Partial}
\end{figure}

Adding Katz-centrality as an explanatory variable has a major impact with more that $10\%$ difference in median probability between highest and lowest central patents (see Figure \ref{fig:Hydrogen_domain_Partial}). Also, we see that adding \textit{Katz}-centrality as a measure reduces the significance of the subdomain effects showing a systematic difference between domains on this variable. This explains the earlier finding that the technological improvement rate is constant \citep[e.g.,][]{benson_quantitative_2015}. As Katz-centrality absorbs the subdomain effect for 'production', 'storage', and 'distribution' it reveals an inherently different knowledge production process. 

As the effect of Katz-centrality is smaller, for example when comparing 'distribution' and 'fuel cells', as can be seen from comparing Figures \ref{subfig:Hydrogen_Subdomains_Distribution_Partial} and \ref{subfig:Hydrogen_Subdomains_Fuel_Cells_Partial}, average Katz-centrality will be lower, because fewer citations will be awarded to higher Katz-central patents. This produces a lower average and is endogenous to each subdomain resulting in a consistent and constant effect, which in turn leads to proportional citation rates. Higher Katz-centrality increases citation speed, but not equally for every subdomain. This shows that in faster developing domains the main branches grow faster, and that more ‘trees’ grow in the forest. This also explains the effect of domain centrality on the technological improvement rate, there is more diverse and more deep organizational learning \citep{march_exploration_1991, cohen_absorptive_1990}.

\

That 'distribution' is the slowest knowledge producing of the $4$ subdomains (production, storage, distribution, and fuel-cells), implies that it is a proportionally increasing cost component over time, because citation structure determines the technological improvement rate \citep{benson_quantitative_2015}. This is an important result as it will support effective business and policy decisions, as we will discuss. In addition to that, another major advantage of shifting the analyses from domain to patent level is that it becomes unambiguous what direction a technology subdomain develops \citep[also see][for an extensive discussion]{park_tracing_2017}. This is of strategic significance to organizations, governments on all levels and society as a whole.

\begin{table}[ht]
\centering
\begin{threeparttable} 
\caption{Cox Proportional Hazards Model Results}\label{tab:CPH_result}
\label{tab:cph_results}
\begin{tabular}{lrrr} 
\toprule
Covariate & Subdomain Model & Knowledge Model \\ 
\midrule
$H_{2}$ Storage & 0.780$^{***}$(0.018) & 1.050$^{*~~}$(0.018) \\
$H_{2}$ Distribution & 0.711$^{~~~}$(0.161) &   \\
$H_{2}$ Production & 0.742$^{***}$(0.017)  & 0.984$^{~~~}$(0.018)  \\
Fuel Cells & 1.046$^{***}$(0.009)  & 1.660$^{***}$(0.018)  \\
Days after pub. &  & 1.576$^{***}$(0.026)  \\
Tot. CPC Class. &  & 1.050$^{***}$(0.004)  \\
Cit. w/in Subdomain &  & 0.651$^{***}$(0.015)  \\
Nodal Katz C. &  & 1.547$^{***}$(0.033)  \\
\midrule
Observations & 1,791,203 & 1,791,203 \\
Citations & 1,314,970 & 1,314,970 \\
\bottomrule
\multicolumn{3}{l}{$^{***} p \leq 0.0001$, $^{**}p \leq 0.001$, $^{*} p \leq 0.01$} \\
\end{tabular}

\begin{tablenotes} 
\small 
\item Results Cox Proportional Hazard Model on Citation Event. Data are standardized to allow direct coefficient comparison. Standard errors are reported in brackets. Liklihood ratio test on difference partial log liklihood between the subdomain Model and Knowledge Model shows highly significant improvement ($p<.00001$). In model 1 $H_{2}$ Distribution is not significant ($p=.034$) implying it can not be distinguished from the baseline hazard. Therefore, we assume the baseline hazard in model 2 is $H_{2}$ Distribution. After including knowledge production variables $H_{2}$ Storage retains significance ($p=.006$, here considered marginal), however $H_{2}$ Production is not significant ($p=.370$) anymore. Much of citation variance in these subdomains is explained by age (publication year), knowledge width (total number of classes in which a patent is classified) and knowledge depth (citation occurs within subdomain), and most importantly by the importance of a patent (Nodal Katz Centrality). This reemphasizes that the differences in the structure of citations due to domain intrinsic characteristics strongly affects knowledge production. As a categorization in Fuel Cells remains informative other unobserved variables may play a role in explaining knowledge production.
\end{tablenotes}
\end{threeparttable}

\end{table}

\begin{sidewaysfigure}
\centering
    \includegraphics[width=\linewidth]{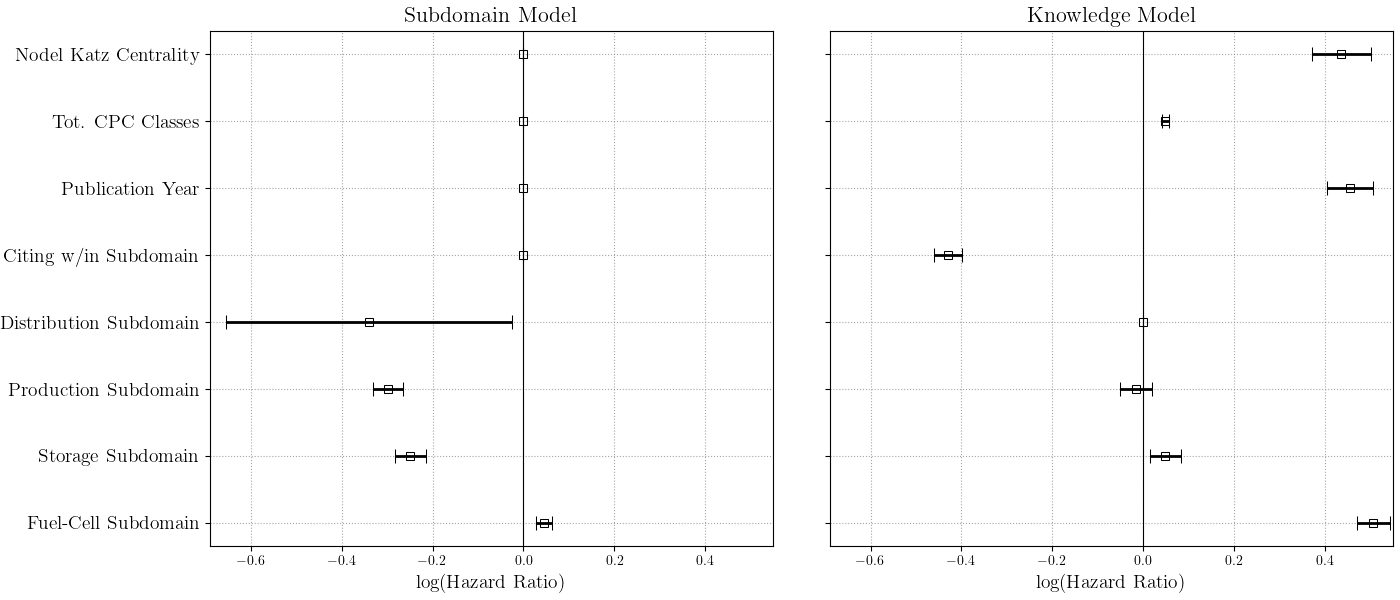}
    \caption{Comparing coefficients between the subdomain and knowledge models. The subdomain model is restricted in the patent attribute and citation variables. The knowledge model restricts the Distribution Subdomain variable. Error bars show 95\%-confidence interval for clarity. At 99\% the Distribution Subdomain is not significant.}
    \label{fig:model_comparison}
\end{sidewaysfigure}
Many of the differences between domains can be explained by the relatively simple model we propose as can be seen from Table \ref{tab:CPH_result} and Figure \ref{fig:model_comparison}. Citations of patents belonging to the same subdomain significantly decrease citation speed. This implies that more domain specific patents on average fall back to older patents, which slows down deep knowledge development or exploitative learning on average. This reveals that exploitative learning is less preferred. Early debates on organizational learning put emphasis on balancing exploratory and exploitative learning from a strategic perspective \citep[e.g.,][]{kogut_knowledge_1992, levinthal_myopia_1993, he_exploration_2004}. It was also contented that especially exploratory learning was often at risk of under investment. Our results suggest that exploitative learning as measured by within subdomain citations results far less often in patenting. This raises the question whether exploitative learning is ignored too often or for example that such developments are kept 'in-house', without patenting. This could also be explained by the fact that profiting from patents is different for broader, enabling technologies and narrower domain specific technologies \citep{gambardella_profiting_2021}.

\section*{Conclusions and Discussion}
Evaluating citation rates per subdomain provides valuable insights into domain level development rates. We can identify bottleneck technologies which in the future will most likely bear the  fastest growing cost component. It is possible to make this statement as the proportionality assumption in the CPH models hold for in distinguishing between subdomains within the larger $H_2$ technology domain. These models are relational event models which estimate the hazard-rate of citation after a previous citation occurred. An extension would be to to identify those patents that drive knowledge development most, if further fine-grained classification is possible. Since, we also identified an endogenous network variable that explains much of the difference in citation rates identifying the relative importance of individual patents over time is also possible. The methodology proposed here thus has great flexibility and provides deep insights into knowledge production, consistent over all levels. This has important consequences for those directly involved in technology development, but also for regulators and policy makers. The knowledge production model allows to identify the rate with and direction which in technology domains evolve, and will contribute to strategic decision-making.  

The substantive results are statistically sound and congruent with the existing knowledge development literature. The finding that older patents get cited less shows that as knowledge accumulates, older knowledge become more obsolete. Alternative explanations could built on institutional arguments, such as the expiration of patents, however this would not explain that some patents are still cited after expiration.

The strong effects of both the breath of cited patents, as well as the effect of deep within domain citations relate to the distinction in knowledge development theory between generalist and specialist knowledge. Although, a direct effect of breadth and depth was not found to affect TIR's \citep{magee_quantitative_2016} here we find that both strongly affect citation rates. It is also interesting to see these difference on average have opposite effects. Deepening of knowledge goes much slower suggesting that technological improvement rates might be much lower than thought before. This could explain variance in the relation between patent network characteristics and productivity gains reported in \citep{magee_quantitative_2016} and shows the effect of a qualitative difference between citations.

The importance of patents contributes to the citation rate, but not equally for every subdomain. This shows that the main branches grow faster, and that more ‘trees’ grow in the forest of more rapidly developing domains. This explains the effect of domain centrality on the technological improvement rate. In this study we can now attribute this to very specific patents. This increases the relevance of looking at development paths as they may indicate the most promising directions for further research, which are a natural target for external support. Not only in terms of governmental, but also support from investors and incumbents. Yet, the embedding of the knowledge production system as revealed in patent citation networks within a larger system may explain why different subdomains develop at different rates.

Most notably in our case, hydrogen distribution will be expected to be a bottleneck in the development of a hydrogen economy as we see that median time to citation is highest in this subdomain. Also, we can infer that this is due to the invention pattern that is typical for hydrogen distribution technological development. As Katz-centrality is lower on average fewer pivotal inventions are published that could fuel the development of the subdomain. We conjecture that this is inherently due to the nature of the distribution technology and it's application in innovations, which will always require substantial economic and social investments. As the investments needed in this subdomain are usually very high there will be less incentive to reveal and more incentive to monopolize new knowledge.

Our findings suggest that cost advantages in production and storage can be achieved more quickly in the short run, while distribution remains a cost bottleneck, resulting in long-term cost dominance and generating equilibrium rents. However, we also highlight the catch-22 situation where these rents are dependent on demand volume but may hinder capacity maximization due to monopolistic optimization.

An interesting question for further research hence becomes: What is the effect of hydrogen distribution ownership on the speed of development? This paper describes, but does not explain the behavior that leads to slower development. Many market forces may play a role in this process. Since large distribution networks are in the hands of large privately owned companies either or both the involvement of these incumbents and governments may be pivotal to succeed the development of a $H_{2}$-economy.

\section*{Acknowledgments}
This work was supported by the UKRI ISCF Industrial Challenge, through the UK Industrial Decarbonisation Research and Innovation Centre (IDRIC) award number: EP/V027050/1, under the Industrial Decarbonisation Challenge (IDC). This paper is an output from IDRIC project 9.3 Knowledge Transfer and Innovation Diffusion. We also thank participants of the "Risk in Decarbonization Finance Series: Venture Capital, Private Equity and Sustainable Investments" (March 9th 2023) held at Panmure House, Edinburgh, U.K., and, those at the IDRIC annual meeting (16-17th May 2023) at the at Kia Oval, London, U.K., for their questions and constructive feedback. Special thanks to David Krackhardt, Matthew Smith, Yasiman Sarabi and Mercedes Maroto-Valer for feedback and unwavering support.


\appendix
    
\subsection*{Lens.org API}\label{appendix:API}

\lstset{language=Python, basicstyle=\ttfamily, keywordstyle=\bfseries, breaklines=true} 

Python code to retrieve data from Lens.org \citep{jefferson_comment_2017}. For replication studies an API-key needs to be obtained from the website.
\begin{lstlisting}
import requests
import json

# This is your API key from Lens.org
api_key = "YOUR\_API\_KEY" 

# Base URL for the Lens.org API
url = "https://api.lens.org/patent/search"

# Define your query parameters
dmnnms = '["Y02E60\\/32", "Y02E60\\/34", "Y02E60\\/36", "Y02E60\\/50"]'
include = '["biblio", "doc_key", "lang"]'
inrecords = []

# Construct the request body 
request_body1 = '''{
    "query": {
        "bool":{
            "must": {
                "terms":  {"class_cpc.symbol": %s}
                },
            "should": [
                {"term": { "cited_by_patent": true } },
                {"term":{ "cites_patent": true } }
            ]
        }
    },
    "sort": [
        {"date_published": "asc"}
    ],
    "size": 100,
    "include": %s,
    "scroll": "1m"
    }''' % (dmnnms, include)

# Make the API request
headers = {
    "Content-Type": "application/json",
    "Authorization": f"Bearer {api_key}"  # Include the API key in the header
}

# Function to handle scrolling through paginated results
def scroll(scroll_id=None, pklname="patent_data.pkl"): 
    global request_body, inrecords # List of previously stored patents, default is empty  
    
    try:
        if not os.path.exists(pklname):
            df_total = pd.DataFrame(columns=['lens_id', 'doc_key', 'lang', 'biblio'])
            df_total.to_pickle(pklname)
        else:
            df_total = pd.read_pickle(pklname)

        if scroll_id is not None:
            request_body = f'{{"scroll_id": "{scroll_id}", "include": {include}}}'

        response = requests.post(url, headers=headers, data=request_body)

        if response.status_code == requests.codes.too_many_requests:
            time.sleep(1)
            return scroll(scroll_id, pklname) 

        elif response.status_code != requests.codes.ok:
            print(f"Request failed with status code: {response.status_code}")
            return response.status_code

        else:
            data = response.json()
            if data.get('results') is not None and data['results'] > 0:
                scroll_id = data['scroll_id']
                df = pd.DataFrame(data['data'])
                # Only add unrecorded patents
                sel = [not id in inrecords for id in df.lens_id.to_list()]
                if any(sel):
                    dft = pd.concat([dft, df[sel]], axis=0) 
                return scroll(scroll_id, pklname) 

        return df_total

    except Exception as e:
        print(f"An error occurred: {e}")
        return None 



response = requests.post(url, headers=headers, data=request_body1)

# Check if the request was successful
if response.status_code == 200:
    # Parse the JSON response
    data = response.json()

    # Process the results (example: print patent titles)
    for patent in data['response']['docs']:
        print(patent['lens_id'], patent['title']) 

else:
    # Print the error message if the request failed
    print(f"Request failed with status code {response.status_code}")
    print(response.text) 
\end{lstlisting}




\bibliographystyle{plainnat}  
\bibliography{Bibliography/patent_citation_speed}

\end{document}